# Contact resistance of various metallisation schemes to superconducting boron doped diamond between 1.9 and 300 K

Scott A. Manifold[1,2,*], Georgina Klemencic[1], Evan L.H. Thomas[1], Soumen Mandal[1], Henry Bland[1], Sean R. Giblin[1], Oliver A. Williams[1]

[1]School of Physics and Astronomy, Cardiff University, Cardiff, UK

[2]EPSRC Centre for Doctoral Training in Diamond Science and Technology, UK

## Abstract

Diamond is a material that offers potential in numerous device applications. In particular, highly boron doped diamond is attractive due to its superconductivity and high Young's Modulus. The fabrication of stable, low resistance, ohmic contacts is essential to ensure proper device function. Previous work has established the efficacy of several methods of forming suitable contacts to diamond at room temperature and above, including carbide forming and carbon soluble metallisation schemes. Herein, the stability of several contact schemes (Ti, Cr, Mo, Ta and Pd) to highly boron doped nanocrystalline diamond was verified down to the cryogenic temperatures with modified Transmission Line Model (TLM) measurements. While all contact schemes remained ohmic, a significant temperature dependency is noted at $T_c$ and at the lowest temperatures the contact resistances ranged from Ti/Pt/Au with $(8.83\pm0.10)\times10^{-4}$ $\Omega$.cm to Ta/Pt/Au with $(8.07\pm0.62)\times10^{-6}$ $\Omega$.cm.

## 1 Introduction

Diamond has attracted attention for various device applications that take advantage of its desirable material properties. These include its high Young's Modulus, wide bandgap and high thermal conductivity [1,2]. In particular, the high Young's Modulus and superconductivity of boron doped nanocrystalline diamond (B-NCD) shows promise in the fabrication of superconducting nanoelectromechanical systems (NEMS) [3]. To ensure high quality factors NEMS require low surface roughness, which means that NCD films need to be polished [4–6]. Such devices need to operate below the critical temperature of the B-NCD, which has been shown to be approximately 4.2 K [7]. Previous work has also shown that the superconducting properties of B-NCD film dominated by its granularity and that it also displays a high upper critical magnetic field of 7 T [8–10].

The fabrication of appropriate electrical contacts is an essential element in the realisation of the design goals of a device. However, the formation of low resistance ohmic contacts to wide band gap semiconductors like diamond is typically non-trivial due to the large work function at the interface to metal contacts, with metal on smooth diamond being subject to a potential barrier of approximately 4 eV [11]. Very low contact resistances are required when dealing with superconducting devices, as any thermal perturbation can significantly affect the device's properties. Any local heating produced at the contact can raise the temperature enough to reduce the critical current density ($J_c$) of the superconductor, and possibly even exceed the critical temperature ($T_c$) [12].

*Corresponding author. E-mail: manifolds@cardiff.ac.uk (Scott Manifold)

It is necessary to consider charge carrier transport through the contact into the material in the normal and superconducting regimes separately. In the case of p-type semiconductor doping, sufficiently high doping concentrations cause the material to undergo a metal-insulator-transition (MIT) where the semiconductor becomes degenerate with its Fermi level entering the valence band [13]. Sufficient doping results in superconductivity at low temperatures. The MIT for diamond occurs at boron doping concentrations of ~$4.5 \times 10^{20}$ cm$^{-3}$ [14]. The formation of Cooper pairs in a superconductor causes a region of suppressed density of states around the fermi energy, resulting in an energy gap. In a junction between a superconductor and a normally conducting material, phase coherent transport can be induced in the normal conductor by the proximity effect [13,15]. In the superconductor, unpaired electrons penetrate causing a reduction in the energy gap and weakening of the superconductivity in the vicinity of the interface. In terms of carrier transport, the system of a metal contact to a highly doped semiconductor that has become superconducting is comprised of three layers. Firstly, the metal itself which is subject to temperature dependent resistivity $\rho(T)$ due to electron scattering where either $\rho \propto T$ or $\rho \propto T^3$ (for transition metals) below the Debye temperature [16,17]. Wieck (1988) showed that this metal resistance dominates as the contact resistance vanishes below $T_c$ for contacts to another superconductor [18]. Next there is an interface layer consisting of the region subject to the proximity effect and defects that interfere with the superconductivity of the doped semiconductor, sometimes including an insulating layer depending on the materials involved. Provided that the insulating portion is 1-10 nm thick it is considered to be "tunnel-thin", but this intermediary layer remains a potentially important consideration due to its influence on the potential barrier of the contact [19]. The final layer is the superconductor itself. Assuming low temperatures and thin potential barriers, transport across semiconductor-superconductor junctions are typically analysed in a framework of a tunnelling mechanism [19]. Electrons from the normal conductor with an energy $E$ (relative to the Fermi energy $E_F$) below the superconducting gap cannot enter the superconductor and instead undergo Andreev reflection [20].

Current transport at normally conducting metal-semiconductor interfaces occurs via either thermionic emission or tunnelling [21,22]. Defects near the contact interface can either narrow the depletion region and increase the probability of tunnelling or lower the effective barrier height. With greater defect or doping concentrations, the tunnelling regime dominates. The reduction in contact resistance associated with annealing and carbide formation is thought to be due to the formation of defects near the metal/diamond interface, which lead to an increase of the tunnelling probability of the carriers [23]. This model of ohmic contacts is further validated by decreases in contact resistance with increasing doping concentrations. Boron doping concentrations in diamond of ~$10^{20}$ cm$^{-3}$ have been shown to maximise this effect [24]. Contacts to diamond with high doping concentrations have been shown to have weak temperature dependence between 30–300 °C [25]. Due to the high boron doping concentrations of superconducting diamond, it could be expected that tunnelling is the dominant carrier transport mechanism and that contact resistance will not increase significantly at very low temperatures.

Several methods have been used to fabricate suitable contacts to diamond including the use of high levels of boron doping and the formation of defect rich layers between diamond and the metal [26]. These defect rich layers have typically involved damage from ion bombardment [23] or the formation of metal carbides by annealing [27,28]. Titanium is commonly used as a carbide forming contact to diamond, partially due to its carbide formation being energetically favourable (see Table 1).

**Table 1: Carbide formation energies for transition metals commonly used for contacts with diamond at 600 °C [29].**

|  | Titanium Carbide | Chromium Carbide | Molybdenum Carbide | Tantalum Carbide |
|---|---|---|---|---|
| ΔG (kJ/mol) | -245 | -65 | -56 | -144 |

The annealing of contacts also offers the benefit of increasing contact adhesion [30]. However, care is required when choosing annealing temperatures and times. Degradation and an increase in contact resistance of Ti/Au contacts has been observed upon annealing at 450 °C due to migration of Ti to the surface of the Au [31]. The addition of Pt has been utilised as a diffusion barrier between the Ti and Au layers. It was noted that the Pt and Au layers interdiffused, but as the Pt barrier was not fully consumed it effectively prevented the Ti contamination at the surface [32,33]. In the case of carbon soluble metals, such as Pd, annealing is thought to increase diffusion and has been shown to significantly reduce contact resistance [25].

As-grown CVD diamond exhibits hydrogen termination, which is conductive. When the hydrogen conductive layer is removed from the surface the resistance of the film increases [34]. Oxygen termination is sometimes achieved via wet chemical treatments, for example with $CrO_3$ or $KNO_3$. This may not influence hydrogen terminated grain boundaries which contribute to surface p-type conduction [35]. Oxygen termination via wet chemical treatment has been shown to only have a minor impact on surface conductivity of boron doped films [36], but it does allow for better adhesion.

Stable ohmic contacts with low contact resistance have been shown on diamond devices at room temperature and for high power applications (see Table 2 below). Work has also been done to pursue contacts lightly doped or intrinsic diamond with high surface conductivity due to hydrogen termination[26,37,38]. Wang et al. (2016) showed that palladium offers the possibility of low resistance contacts without carbide formation to single crystal undoped diamond. Also of note is the recent work by Xing et al. (2020) which shows the stability of palladium contacts to hydrogen terminated Type IIa single crystal diamond between 300 K and 4 K, with contact resistance increasing from $(8.4\pm1) \times 10^{-4}$ $\Omega cm^2$ to $(1.3\pm0.2) \times 10^{-3}$ $\Omega cm^2$ in this temperature range.

**Table 2: Comparing contact resistance ($\Omega cm^2$) of different metallisation schemes.**

|                  | Ti/Au              | Cr/Au       | Mo/Au              | Ta/Au              | Pd                  |
|------------------|--------------------|-------------|--------------------|--------------------|---------------------|
| *Hoff 1996*      | 8.1 x $10^{-2}$    |             |                    |                    |                     |
| *Hewett 1993*    | 3.2 x $10^{-6}$    |             | 1.2 x $10^{-3}$    |                    |                     |
| *Nakanishi 1994* | ~$10^{-6}$         |             | ~$10^{-6}$         |                    |                     |
| *Venkatesan 1993*| ~$10^{-5}$         |             |                    |                    |                     |
| *Chen 2004*      | ~$10^{-4}$         |             |                    |                    |                     |
| *Yokoba 1997*    | ~$10^{-5}$         | ~$10^{-5}$  | ~$10^{-5}$         |                    | ~$10^{-5}$          |
| *Fang 1989*      |                    |             |                    | 1 x $10^{-3}$      |                     |
| *Zhen 2002*      |                    |             |                    | 7.19 x $10^{-5}$   |                     |
| *Wang 2015*      |                    |             |                    |                    | 4.93 x $10^{-7}$    |
| *Xing 2020*      |                    |             |                    |                    | 8.4 x $10^{-4}$     |

Note: Interlayers of Pt as a diffusion barrier are present in some of the above. Where multiple values are provided, the lowest are quoted.

This work seeks to test the efficacy of both carbide-based contacts and palladium contacts on B-NCD down to the cryogenic temperatures necessary for superconducting devices to function.

## 2 Experimental methods

### 2.1 Modified TLM measurements

Measurements of the contact resistance were conducted via a linear Transmission Line Model (TLM) pattern, as is typical for contacts to semiconductors and indeed for diamond at room temperature and above [39]. Contact resistances with superconducting materials are often measured with simpler two or four-probe methods, because the effect of sheet resistance is removed [18,40].

In standard TLM measurements it is critical that the contact pads are formed on a mesa, as this avoids lateral current crowding at the contact and an overestimation of the contact resistance [41]. For a four-wire measurement to two contact pads on a semiconducting mesa, the total measured resistance ($R_T$) will be a combination of the metal resistance ($R_M$), the sheet resistance ($R_s$) and the contact resistance ($R_c$) as follows (see also Fig. 1b):

$$R_T = 2R_C + 2R_M + R_S \quad (1)$$

Acquiring $R_T$ from the measurement of several sets of contact pads with different separations, the contact resistance can be derived from a linear fit of these points. At $x = 0$, $y = 2R_c$ and at $y = 0$, $x = -2L_T$. With standard TLM, $L_T$ is the effective length of the contact due to current crowding. However, because this calculation requires a linear extrapolation to the x-axis intercept assuming constant sheet resistance, the potential for carbide formation at the diamond surface requires the exclusion of this parameter from the TLM calculations. Additionally, in the superconducting regime the concept of such a linear extrapolation, with a gradient $R_S/W$, becomes nonsensical.

The contact resistivity ($\rho_C$) can be extracted from the y-intercept and the contact width ($W$) via the following relations:

$\rho_C = R_C W L_T$ (2)

$\rho_C = R_C W$ (3)

## 2.2 Sample preparation

The growth substrate of high resistivity silicon wafer buffered with 500 nm of $SiO_2$ was seeded with a monodisperse aqueous colloid of ~5 nm diamond nanoparticles [42]. B-NCD film was grown via microwave assisted CVD with the use of low $CH_4/H_2$ chemistry (<3% $CH_4$) [43]. The addition of trimethylboron provided a gas phase B/C ratio 12800 ppm. By calculating directly from this B/C ratio and comparing with earlier works it can be assumed that the boron concentration in the samples is greater than $2 \times 10^{21}$ $cm^{-3}$ [44–46]. This doping concentration provides reliable superconducting performance and ensures that the dopant dependent depletion region at the contacts is minimised. The Raman shift of the film was measured with a Horiba LabRAM HR Evolution equipped with SynapsePlus Back-Illuminated Deep Depletion (BIDD) CCD (see Fig. 1). Laser wavelengths of 473nm, 532nm and 660nm were used. Analysis of these spectra shows the absence of the typical diamond peak (D) at 1332 $cm^{-1}$, with it red shifted into the Fano-like shoulder ($D_F$) at 1285 $cm^{-1}$. Previous work shows that the absence of the D peak is an indication of SIMS measured B concentrations of at least $10^{21}$ $cm^{-3}$ [47–49]. A strong peak (B) at ~1220 $cm^{-1}$ is also observed, attributed by Sidorov and Ekimov to carbon-carbon bonding states where the presence of boron leads to local distortions to the lattice structure [50]. The intensity of this peak correlates with the doping concentration [48,49].

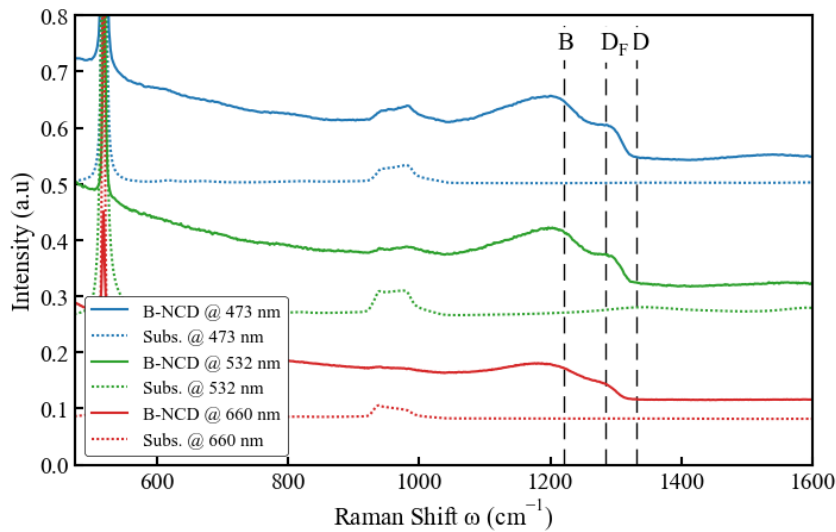

**Figure 1:** Three wavelength Raman shifts of the B-NCD film and substrate (donated "B-NCD" and "Subs."), where the lack of diamond peak (D) at 1332 $cm^{-1}$ and strong peak (B) at ~1220 $cm^{-1}$ is indicative of high levels of boron doping.

The film thickness was 200 nm and $T_c$ was determined to be approximately 2.4 K via a silver epoxy bonded Van Der Pauw pattern. Superconducting and normal state properties of the film are consistent with others produced under the same conditions [10], and also comparable to those produced in the same system over several years of operation (see Fig. 2).

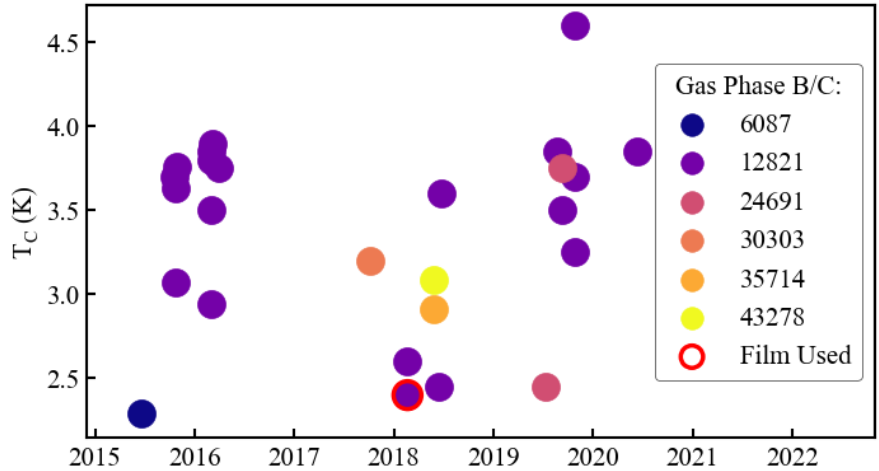

**Figure 2:** The critical temperature of BDD films grown on different dates with various gas phase B/C ratios. The film growths are tuned for different purposes and have various thicknesses but show consistent superconducting performance.

The wafer was diced and prepared into mesa patterns (200 x 1200 µm) via photolithography to produce a nickel mask and subsequent etching by ICP-RIE. The mesas were then ashed in an oxygen plasma (1 min at a power of 30 W with 30 SCCM $O_2$ at a pressure of 0.1 mT) to achieve oxygen termination.

Next, the metal contacts were created via photolithography and magnetron PVD sputtering. The contacts were 200×100 µm, with separations of 160, 80, 40, 20, 10 and 5 µm, giving a range of measurable separations between 5 and 815 µm. Of the five contact schemes prepared, four involved a carbide forming layers (Ti, Mo, Cr, Ta) and one did not (Pd). The carbide forming schemes were all deposited as a trilayer with Pt and Au, with thicknesses of 50 nm for both the interface metal and Pt layers, and capped with 50 nm of Au in a magnetron PVD system without breaking vacuum. To increase the yield of the contacts, before lift-off the samples were topped up with 150 nm of Au in a thermal evaporation system for a total contact thickness of 300 nm. The Pd contacts were 200 nm, limited by the cost of the material.

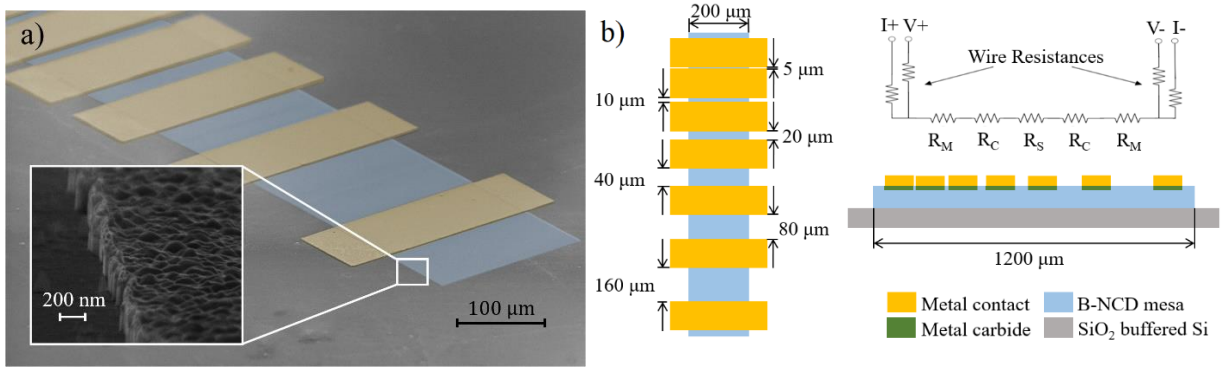

**Figure 3:** (a) false colour tilted SEM of metal contacts on a B-NCD diamond mesa on $SiO_2$ buffered silicon, with zoomed inset showing the edge profile resulting from the ICP-RIE process. (b) plan (left) and cross section (right) diagrams of the TLM patterns used.

To increase adhesion and to create a low resistance ohmic contacts, various annealing protocols were investigated for titanium [23,27,51–55], chromium [25], molybdenum [24,25,30], tantalum [56,57] and palladium [25,58]. Annealing parameters of 600˚C for 10 mins for the carbide forming schemes and 400˚C for 3 mins for palladium were chosen as optimal from the reviewed literature. For this work, annealing was carried out in a rapid thermal annealer after a pump/purge cycle with nitrogen.

Samples were taken through a temperature range of 1.9-300 K in a Quantum Design Physical Properties Measurement System (PPMS), and measurements were taken with custom external electronics including a Picowatt AVS-47B Resistance Bridge. Contact pads were wire bonded with 1% Si/Al 12 µm wire to a carrier chip, and then four-point wired to the PPMS sample puck. Some adhesion variability was encountered when wire bonding the contact pads, so resistance measurements of ten contact separations were measured to maximise accuracy.

## 3    Results and discussion

Resistance measurements between pads of different separations were carried out for each metal contact scheme. Above the superconducting transition, the total resistance of each scheme scales with temperature, with the curve following the expected $T^{1/2}$ dependence [10]. These were then combined to show the dependence of the resistance on separation distance (see Fig. 4(a)). Isothermal slices of these data were then taken, allowing modified TLM calculations to be carried out at each temperature value using a linear least squares fit (see Fig. 4(b)). The fitting procedure also provides error values. As discussed previously the $L_T$ term was excluded from the main calculations. The results are shown in Fig. 5 and summarised in Table 3. During process and adhesion optimisation, measurements were performed on samples taken from a separate wafer grown under the same conditions as the primary film. Despite being limited to a few data points due to bond failure, the Ti based scheme provided contact resistance values of 1.7 x$10^{-3}$ Ω.cm showing reasonable agreement to the final data. To allow comparison to literature values, conventional TLM calculations (with $L_T$ included) were performed at 300 K. The values at 300 K show contact resistivity values in the region of ~$10^{-7}$ Ω$cm^2$, which compares favourably to those found in literature.

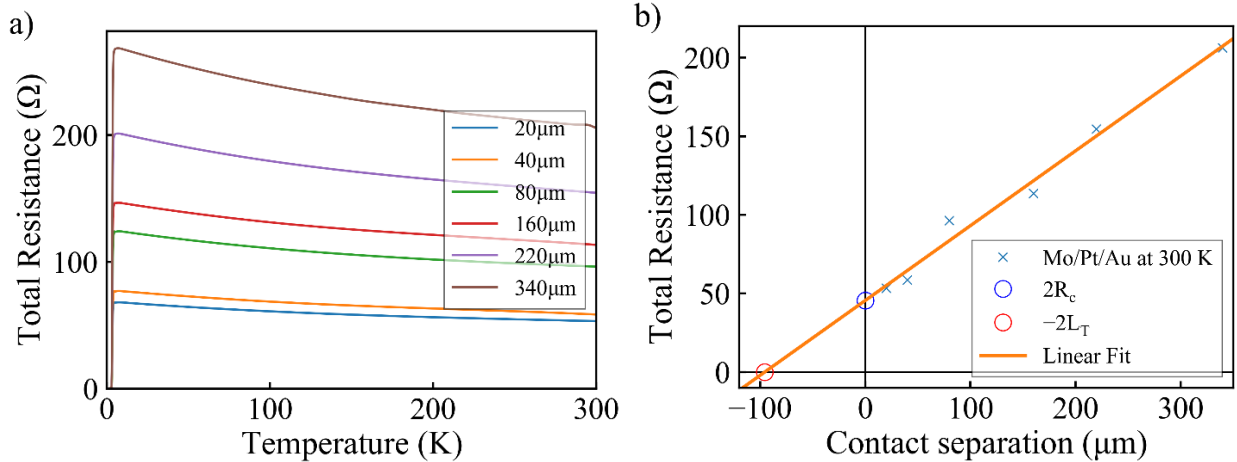

**Figure 4:** (a) example dataset of contact resistances of different contact separations across the temperature range, (b) example linear fit of an isothermal slice through the data (error bars derived from precision of these single resistance measurements are of negligible size).

As typical TLM methods rely on fitting total measured resistance with contact separations, the extrapolation to the contact resistance value includes the sheet resistance of the diamond, which gives rise to the gradient of the fitted line. As such, when the sheet resistance is zero in the superconducting regime, this line becomes flat at $y = 2R_M + 2R_c$ and the measurements effectively approximate typical superconducting two and four probe contact resistance measurement methods. However, it was found that $R_T$ in the superconducting regime was relatively high for all contact schemes, invalidating the possibility of extracting the contact resistance from the four probe measurements. For this reason, TLM calculations were also used in the superconducting regime.

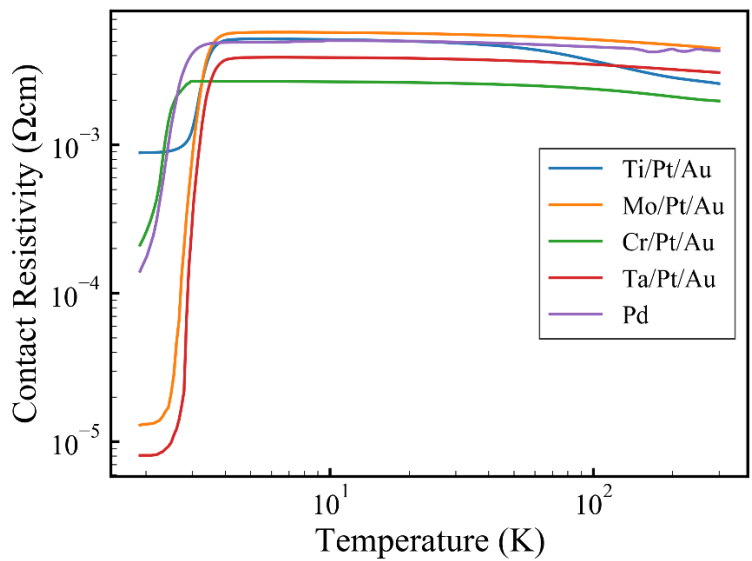

**Figure 5:** Contact resistivity of the various contact schemes scaling with temperature. From 300K to the superconducting transition all the schemes show a slight increase in resistivity, where a significant reduction of the contact resistivity is then observed.

The contact resistance of all metallisation schemes show good stability and minimal signs of a temperature dependence between 300 K and the superconducting transition. A considerable decrease in contact resistance in all samples was noted at approximately 2-4 K, with the variation in the temperature of this decrease attributed to variations in the critical temperature across the film from which the samples were fabricated. All contacts appear to follow approximately the same dependency until this point. This is in contrast to the results obtained by Xing et al. (2020), where carbide forming Ti based contacts to undoped single crystal diamond became non-ohmic at low temperatures while Pd contacts remained stable. This is presumably due to high doping levels of superconducting B-NCD allowing greater tunnelling efficiency which eclipses thermionic emission.

**Table 3: Contact resistance of the tested metallization schemes at 300 K and 1.9 K, with and without the effective contact length factor $L_T$ respectively.**

| Contact Resistance | Ti/Pt/Au | Cr/Pt/Au | Mo/Pt/Au | Ta/Pt/Au | Pd |
|---|---|---|---|---|---|
| At 300 K (incl. $L_T$) (x$10^{-7}$ $\Omega$.cm$^2$) | 2.55±0.1 | 0.82±0.1 | 8.03±0.3 | 3.81±0.4 | 4.99±0.3 |
| At 300 K (excl. $L_T$) (x$10^{-3}$ $\Omega$.cm) | 2.58±0.15 | 1.97±0.12 | 4.46±0.29 | 3.06±0.28 | 4.48±0.35 |
| At 1.9 K (excl. $L_T$) (x$10^{-4}$ $\Omega$.cm) | 8.83±0.10 | 2.11±0.04 | 0.13±0.01 | 0.08±0.01 | 1.40±0.17 |

Between 300 K and the superconducting transition, the Ti based contacts compare favourably with the other schemes. In the superconducting regime the Ti contacts did not show the same magnitude of decrease in contact resistance as the other metals, resulting in the highest contact resistance at 1.9 K. The ranking of the contact resistance of the metals is also different between the normal and superconducting regimes. This ordering does not show direct correlation with the carbide formation energies in Table 1. However, given the values in Table 1 and that all carbide forming metal contacts were annealed with the same parameters, different thicknesses of each carbide might be present in the interface.

In contrast to Wieck (1988), a vanishing contact resistance below $T_c$ was not broadly observed, although this behaviour may be masked by the lower $T_c$ in the Cr and Pd contacts. This is perhaps due to the formation of the carbide layers or other defects in the interface layer increasing the height of the potential barrier, width of the insulting layer or size of the region subject to the proximity effect. Of note is the significantly higher resistivity of titanium carbide compared to the other carbides (Table 4), which adds some credence to this theory and perhaps explains the higher contact resistance of the Ti contact in the superconducting regime. Furthermore, this variability in the resistive contribution of the carbide layer justifies the exclusion of the $L_T$ term from the TLM calculations.

**Table 4: Electrical resistivity of metal carbides at 20 °C.**

|  | Titanium Carbide | Chromium Carbide | Molybdenum Carbide | Tantalum Carbide |
|---|---|---|---|---|
| $\rho$ ($\Omega$m) | 3-8 x $10^{-3}$ [59] | 1.47 x $10^{-8}$ [60] | 9.7 x $10^{-7}$ [59] | 3.6 x $10^{-7}$ [61] |

In the course testing each pad separation, all samples were taken through numerous temperature cycles between 300 K and 1.9 K without any degradation. While it was occasionally noted that wire bonds would separate from contact pads, this was probably due to mechanical tension necessarily introduced during wire bonding to meet the demands of the layout. Failure of the wire bonds was also noted at the contact pads of the carrier chip and PPMS measurement puck.

## 4 Conclusion

In terms of the implication to device applications, all the metallisation schemes tested have approximately equivalent contact resistivity (and therefore local heating) at room temperature and down to the superconducting transition, but differences emerge when the substrate is superconducting. In this regime, the titanium scheme performs the least favourably out of the five tested with the molybdenum and tantalum schemes providing the lowest contact resistance. It can therefore be stated that carbide forming and carbon soluble metallisation schemes allow fabrication of suitable contacts to superconducting diamond devices, but some consideration should be given to the use of Ta over Ti interfaces.

It has also been shown that the high doping concentration of superconducting B-NCD ($\sim 10^{20}$) preserves the Ohmic nature of carbide forming contacts at low temperatures, whereas contacts to intrinsic diamond with low temperature requirements are perhaps limited to other options.

The mechanical stability of the tested carbide and carbon soluble contacts has also been verified across the temperature range.

## 5 Acknowledgements

The authors would like to acknowledge M. Salman and J. A. Cuenca for many useful discussions. We also gratefully acknowledge support by the European Research Council under the EU Consolidator Grant "SUPERNEMS" (Project No. 647471), and funding received from the EPSRC Centre for Doctoral Training in Diamond Science and Technology, United Kingdom Grant No. EP/L015315/1.